# Anti-Counterfeiting for Polymer Banknotes Based on Polymer Substrate Fingerprinting

Shen Wang, Ehsan Toreini, and Feng Hao, *Senior member, IEEE*

***Abstract*—Polymer banknotes are the trend for printed currency and have been adopted by more than fifty countries worldwide. However, over the past years, the quantity of polymer counterfeits has been increasing, so has the quality of counterfeits. This shows that the initial advantage of bringing a new polymer technology to fight against counterfeiting is reducing. To maintain one step ahead of counterfeiters, we propose a novel anti-counterfeiting technique called Polymer Substrate Fingerprinting (PSF). Our technique is built based on the observation that the opacity coating, a critical step during the production of polymer notes, is a stochastic manufacturing process, leaving uneven thickness in the coating layer and the random dispersion of impurities from the ink. The imperfections in the coating layer result in random translucent patterns when a polymer banknote is back-lit by a light source. We show these patterns can be reliably captured by a commodity negative-film scanner and processed into a compact fingerprint to uniquely identify each banknote. Using an extensive dataset of 6,200 sample images collected from 340 UK banknotes, we show that our method can reliably authenticate banknotes, and is robust against rough daily handling of banknotes. Furthermore, we show the extracted fingerprints contain around 900 bits of entropy, which makes it extremely scalable to identify every polymer note circulated globally. As compared with previous or existing anti-counterfeiting mechanisms for banknotes, our method has a distinctive advantage: it ensures that even in the extreme case when counterfeiters have procured the same printing equipment and ink as used by a legitimate government, counterfeiting banknotes remains infeasible because of the difficulty to replicate a stochastic manufacturing process.**

## I. Introduction

Despite the increasing volume of transactions made by credit cards and electronic payment methods, banknotes still play a crucial role in our society. In many countries, such as the US, the UK, Canada, Australia, and the European Union, the demand for cash continues to grow with the value of banknotes in circulation increasing each year typically by a factor of 5 to 10 percent [1]. Globally, there are over 500 billion banknotes in circulation. According to a report by McKinsey & Company [2], over the past years, although the share of the world's transactions carried out in cash has fallen, banknotes remain one of the most widely used payment instruments in the world.

Counterfeiting, or the forgery of banknotes, has been a major threat to the society and economy. Since most banknotes cost little to produce, a successful forgery is virtually all profit. People who fall victim to this crime are essentially robbed. Their losses cannot be reimbursed as doing so will facilitate the circulation of counterfeits and encourage illegal activities. Widespread counterfeiting can severely undermine the value of the currency, and disrupt the economic development [3].

In general, anti-counterfeiting methods challenge the forger in two main aspects: the substrate, and the printing. Traditional banknotes use a paper substrate made of cotton and linen. Compared with the bond paper made of wooden particles, the cotton/linen paper is substantially more expensive and more durable. When used for banknotes, it also contains various security features which are introduced during the manufacturing process, such as watermark, embossed metallic thread and other unique features. The printing is another aspect that gives banks an edge against counterfeiting. It requires specialised equipment and ink which are prohibitively expensive for counterfeiters. One of the most important printing techniques is the so-called intaglio (gravure) printing, which gives the raised print and the unique texture feel of a banknote [3].

The latest development in banknotes is to print them on polymer: a thin, flexible plastic [3]. The new polymer substrate not only supports traditional security printing as employed for paper notes, but also allows enhanced security features, such as see-through window and foil patch. This makes them harder to counterfeit than paper notes. Since the first introduction in Australia in 1988, they have become the trend for printed currency and have been adopted by more than fifty countries. In the UK, Bank of England first issued polymer £5 and £10 in 2016 and 2017, respectively. It has started replacing £20 with polymer notes since 2020.

The introduction of polymer banknotes has evidently reduced counterfeiting. For example, after Australia fully replaced paper banknotes with polymer series in 1996, the rate of counterfeits fell noticeably from 16 ppm (parts per million - the number of counterfeits per million genuine banknotes in circulation) in 1996 to only 3 ppm in 2000 [4].

However, counterfeiters have been catching up. After 2000, the counterfeiting rate in Australia gradually increased, and reached above 25 ppm in 2015 [4]. As the quantity of counterfeits increases, so does the quality. While the first recorded counterfeits on polymer were detected in 1997, they were printed on a paper substrate and used techniques only to simulate the feel of polymer. Around 2010, polymer counterfeits began to appear by using advanced technologies that enabled counterfeiters to print large volumes of counterfeits on a plastic film. This shows that the initial advantage of bringing a new polymer technology to fight against counterfeiting is

Shin Wang and Feng Hao are with Department of Computer Science, University of Warwick, Coventry, CV4 7AL, U.K. E-mail: shin.wang@warwick.ac.uk; feng.hao@warwick.ac.uk. Ehsan Toreini is with Department of Computer Science, Durham University, Durham DR1 2HM, U.K. E-mail: ehsan.toreini@durham.ac.uk

reducing.

Although polymer banknotes have many existing anti-counterfeiting features, one fundamental limitation for the security assurance of those features is that they critically rely on the difficulty for counterfeiters to obtain the same or equivalent printing equipment and ink. As shown by the example of [5], professional counterfeiters often exploit weaknesses in the supply chain for the manufacturing of banknotes and obtain from worldwide suppliers essentially the same or equivalent printing equipment and ink as used for printing genuine notes. Their chance of success can be significantly boosted when the operation is backed by a state government. For example, many high-quality counterfeits of the US$100 bill, known as “superdollars”, are allegedly made by countries that are antagonistic toward the USA. Some of the counterfeits are of such high quality that, according to Europol, they “are just U.S. dollars not made by the U.S. government” [6]. In face of such professional counterfeiters backed by a state government, existing security features of a banknote can be easily bypassed.

To maintain one step ahead of forgers, we propose a new anti-counterfeiting technique called Polymer Substrate Fingerprinting (PSF). In contrast to existing banknote security features which require delicate design and printing, our technique exploits the stochastic nature of the polymer substrate manufacturing process. It works by analysing the random translucent patterns of the polymer substrate when it is back-lit. These patterns are caused by stochastic printing and the randomly dispersed impurities in the ink during the opacity coating procedure. They naturally occur during the banknote production, and cannot be precisely controlled or duplicated. We show these patterns can be reliably captured by a commodity film scanner and processed into a compact fingerprint to uniquely and reliably identify each banknote.

Our contributions are summarised as follows. First, we propose Polymer Substrate Fingerprinting (PSF), a novel anti-counterfeiting technique for polymer banknotes based on analysing the naturally occurring, unique, and unrepeatable imperfections in the opacity coating layer of a polymer substrate. Second, we present a proof-of-concept implementation that uses a commodity negative-film scanner to capture those imperfections by photographing the random translucent patterns of a polymer substrate when it is back-lit and transforming them into a compact fingerprint for authentication. Third, we collect an extensive dataset using the UK polymer banknotes and conduct experiments to show that our technique can reliably authenticate banknotes with high accuracy, is robust again rough daily handling, and is highly scalable to identity every polymer banknote circulated in the world.

## II. Production of Polymer Banknote

While the world’s first banknote printed on clear plastic film was issued in Australia in 1988, this was the result of nearly twenty years of research and development. The major breakthrough in the field was the invention of a special type of plastic called biaxially-oriented polypropylene (BOPP), which after being covered with opacity coating allows quality printing of all of the security features that are printed on traditional paper notes [7]. The use of BOPP makes the polymer banknote highly durable, as well as being waterproof and dirt-resistant.

A polymer note starts as clear plastic heads, which are melted down at a high temperature (around 166 °C) and then blown into a large bubble of several storeys high. The walls of the bubble are pressed together and cooled to form a laminated polymer film. A layer of opacity coating will be added to allow printing security features on the polymer film.

The opacity coating process applies white ink to the film to make it opaque, except for areas that are left clear as see-through windows. The see-through window is a security feature applied on the polymer note as it forces a forger to use clear plastic film as the substrate, which requires more advanced printing equipment than a paper substrate.

The technique used for opacity coating is called *gravure* printing. Figure 1 shows an overview of the process. The substrate is pressed against the inked cylinder on a rotary press between a backing roller and a gravure roller. The cylinder is etched with small cells on the edge which hold the ink fetched from a liquid pool. When the cylinder is partially immersed in the liquid pool, it picks up ink to fill its recessed cells on each rotation of the press. A flexible blade (also known as the “doctor blade”) is used to remove any excess ink from the printing cylinder, leaving ink only in the cells.

At a microscopic view, the opaque ink layer after the gravure printing process is highly non-uniform, showing random variations in the thickness, as shown in Figure 1(b). This is due to two main reasons. The first is related to air bubbles. When the ink in the cell is transferred to the substrate under the pressed contact, air meniscuses penetrate the gap and become air bubbles trapped in the ink [8]. Due to the air bubbles, the ink transferring process is only partially performed. The second reason is related to the solid residues. After the ink is transferred to the substrate, the remaining liquid in the cell evaporates, leaving a solid substance. The substance adhering to the bottom of the cell reduces the volume of the container. As a result of a combined effect of air bubbles and solid residues, the opaque link layer is highly uneven. The uneven coating layer causes the polymer substrate to exhibit random translucent patterns when it is back-lit by a light source, which we will demonstrate later. The existence of impurities in the ink adds further randomness to these patterns. All these are the imperfections from the opacity coating process, and they constitute the physical basis for the anti-counterfeiting technique that we propose in this paper.

After the white ink coating, the polymer substrate is ready for the subsequent printing of security features. Our technique does not rely on any of the printed security features, however we describe the process here for completeness. Security printing involves several layers of printing applied in sequence. The first is *offset litho*, which uses an offset roller to transfer ink to the polymer substrate and puts the basic pattern of the banknote in place. This is followed by *intaglio printing*, which is used to put the major design elements such as the portrait and narrative elements (e.g., Her Majesty the Queen on a £10 note). The next is *letterpress*, which prints letter and digits including the unique serial number. The subsequent stage is to print special line patterns on a polymer substrate to form

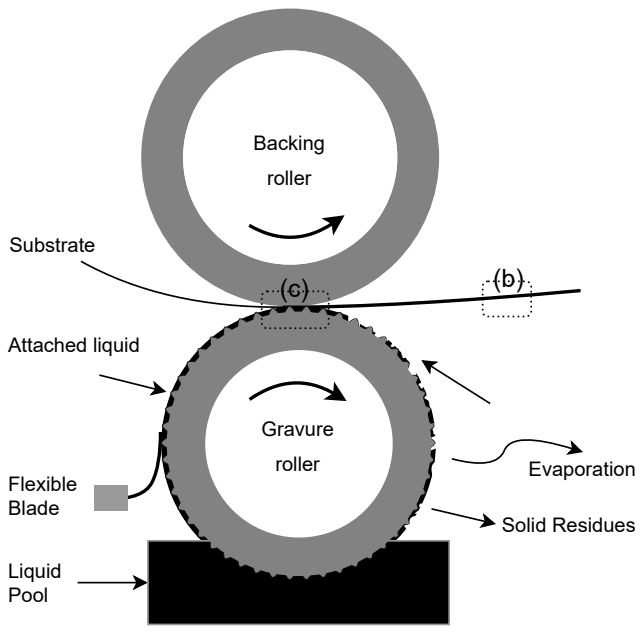

(a) The roller

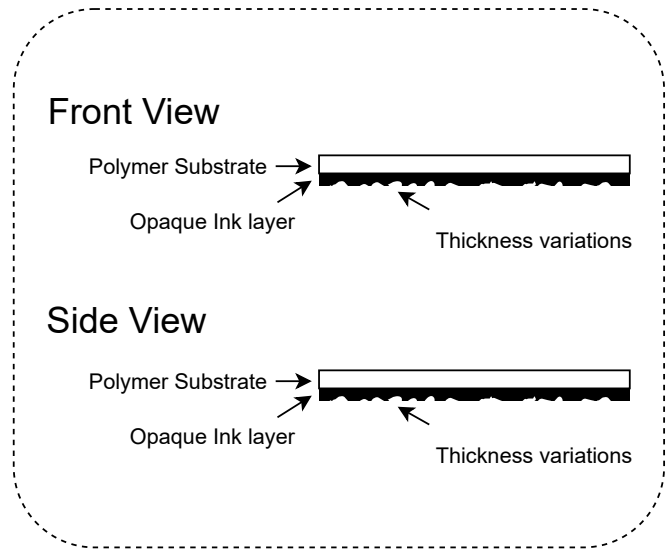

(b) The front and side view

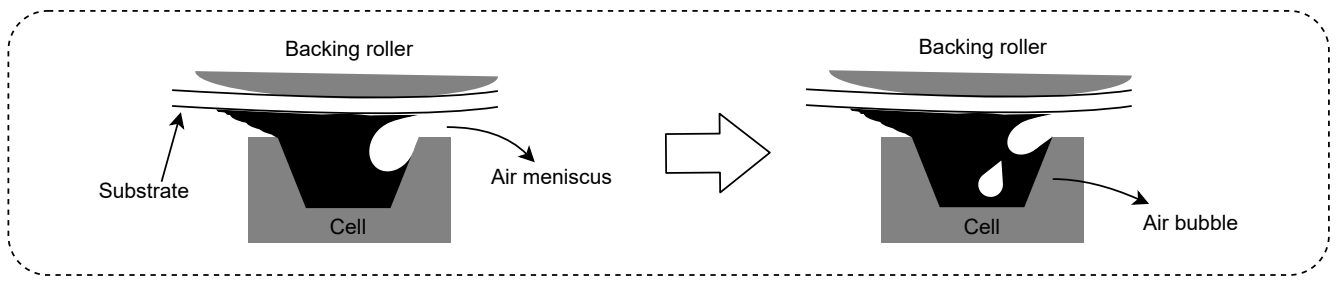

(c) An example of a bubble developing in the cell

Fig. 1: Schematic of the Gravure printing process

diffraction gratings, which typically consist of 12,000 lines per centimetre coated with a thin film of a reflecting metal (e.g., aluminium). Light is diffracted from the lines to give changing colours when viewed from different angles. Next, a protective over-coating ink (clear varnish) is applied on both sides of the note to protect the printed design from dirt and solvent. The tactile features are then applied to assist the visually impaired to identify different denominations. Finally, the printed sheets are guillotined into individual banknotes. Each banknote is then electronically inspected to ensure their quality fulfils the required standard. More details about the polymer note production can be found in [7].

## III. Proposed Solution

### A. Feature Area

First of all, we need to identify an area on the polymer banknote for feature extraction. Based on the observation that the opacity coating is a stochastic process, the ideal areas for feature extraction are those that are directly exposed from the opacity coating and not obstructed by the subsequent security printing. Therefore, for £10 notes, we choose an area between the "Ten" hologram and the see-through window as shown in Figure 2 (a). To locate the area precisely, we use two auxiliary markers: the pound sign in the see-through window and the silver foil patch contained in the hologram. Both are metallic images made by diffraction grating printing at extremely high precision (around 12,000 lines of thin metal film coated per centimetre). These images are darker than the surroundings. Hence, they can be easily separated from the background. Based on the detected markers, the feature area is automatically located with the same position and dimension. Figure 2 (b) displays the snapshots of the same feature area from three different polymer £10 notes when they are back-lit by a light source. These pictures exhibit random translucent patterns, which we will process later. Similarly, we identify and locate the feature areas on a polymer £5 note and a paper £20 note as shown in Figure 2 (c) and

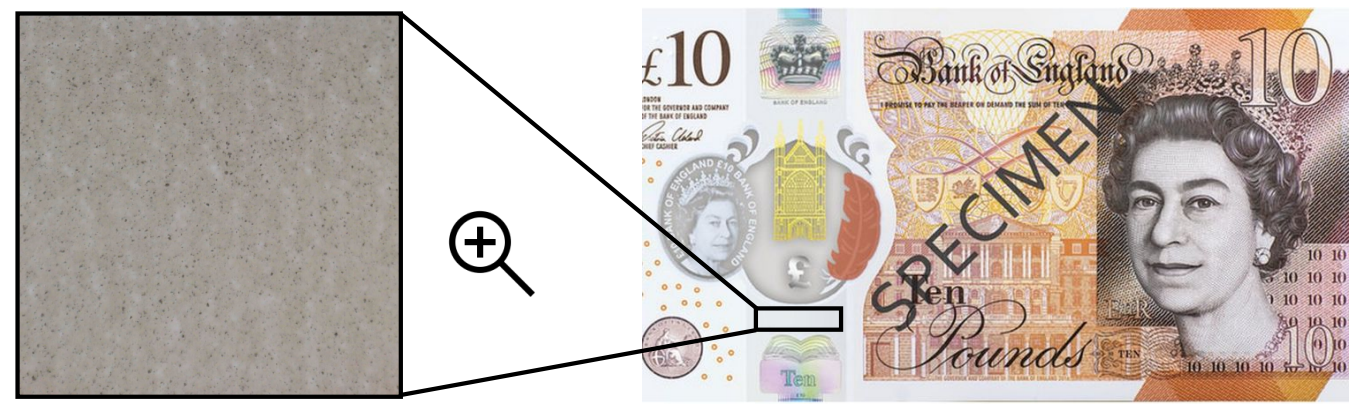

(a) Feature area on £10 note ($19.3\,mm \times 5\,mm$)

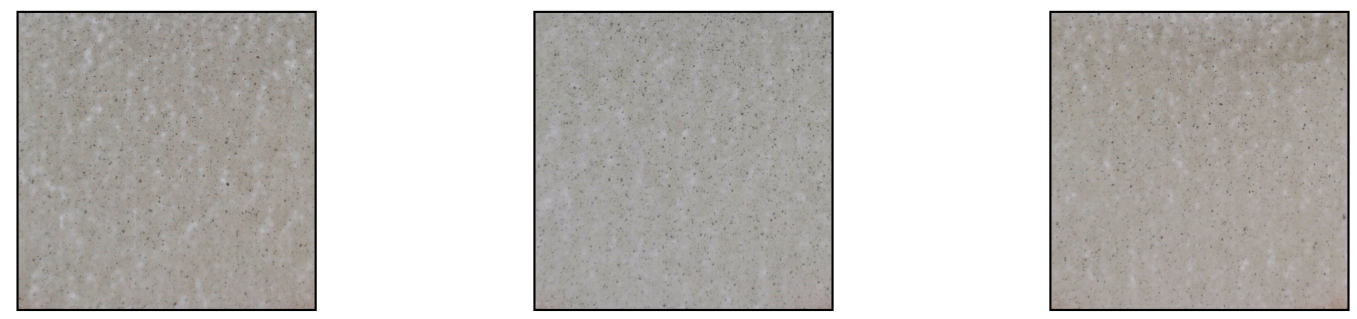

(b) Feature areas on different £10 notes

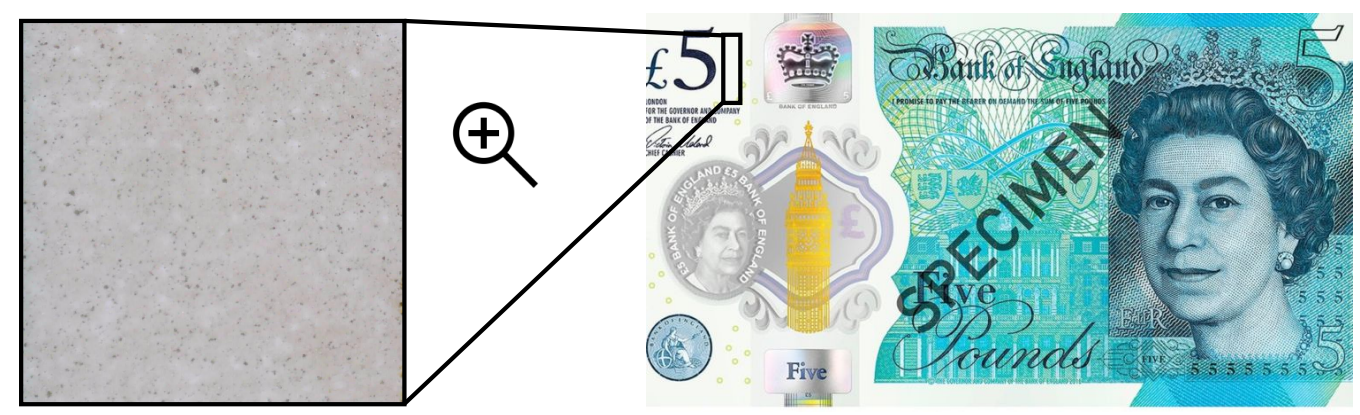

(c) Feature area on £5 note ($3.3\,mm \times 13\,mm$)

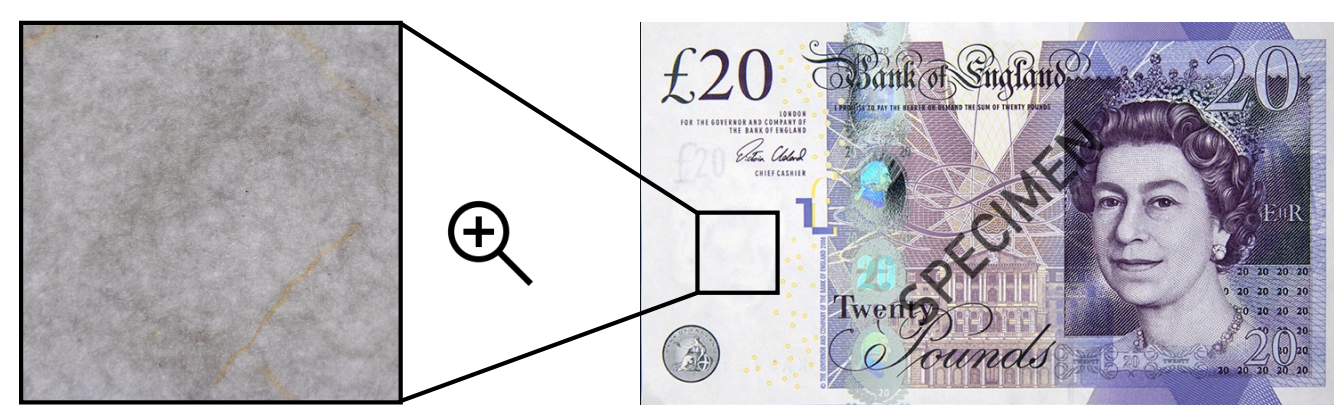

(d) Feature area on £20 note ($16.2\,mm \times 16.2mm$)

Fig. 2: Feature extraction on different banknotes (the zoomed-in pictures are cropped as a square from the original images for demonstration)

(d), respectively. Here we choose the paper £20 note as an example for comparison. When back-lit, a paper banknote also shows translucent patterns, but they are caused by the random interleaving of the cotton linens rather than the uneven coating as seen in a polymer note. Although we focus on the anti-counterfeiting for polymer notes, our technique can also be applied to prevent forgery of paper notes. The performance of our fingerprinting technique for these two different substrates will be compared in the evaluation section.

### B. Experiment Setup

To capture the random translucent patterns of the polymer substrate when it is back-lit, we choose an off-the-shelf negative film scanner (Epson V850), as shown in Figure 3. The resolution of the scanner is set to 3200 dpi to obtain high-resolution images with the help of an embedded back-light. In our experiments, we use a film-frame to hold the banknote. The frame helps to keep the banknote flat and in position during the scanning process.

The primary reason for using a negative-film scanner instead of a more common flatbed scanner is that the former is specifically designed to scan a film by shining light through it, while using light sensors to capture the image on the other side. This fits precisely our purpose. On the contrary, a flatbed scanner scans an object using reflective light. In the UK and other countries, it is prohibited to scan a banknote in this way as it may allow a casual counterfeiter to produce a fake copy. In fact, the firmware of a flatbed scanner has an embedded function to search for anti-copy patterns, e.g., EURion constellation [9] printed on banknotes. Once the scanner finds such patterns, it will stop the scanning process. By contrast, with the film scanner, when the light shines through the £10 banknote, the EURion pattern has been blended into the background. As a result, the obtained image is extremely "noisy", and totally unsuitable for counterfeiting. On the other hand, the "noise" or the randomness in the image is exactly what we need for building an *anti-counterfeiting* system.

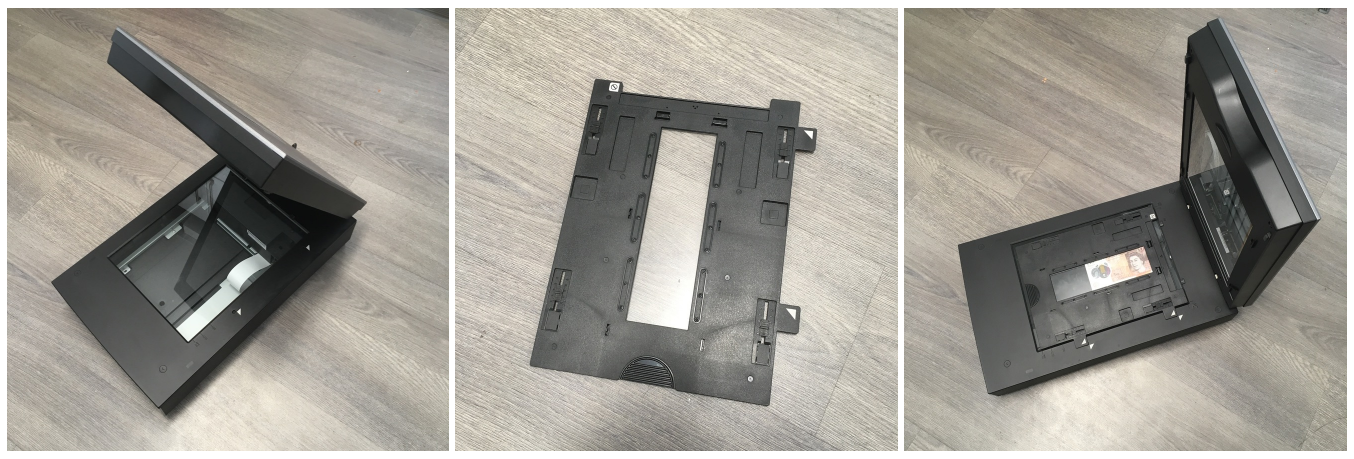

(a) Scanner (b) Modified frame (c) Scanner set

Fig. 3: Scanner setup

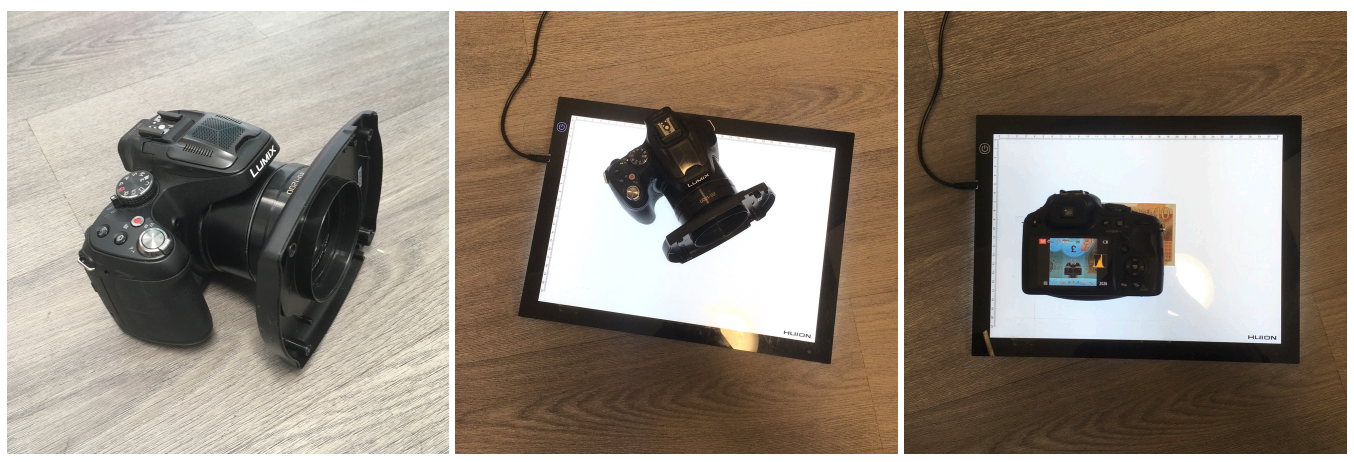

(a) Camera (b) Camera/Light-box (c) Camera set

Fig. 4: Camera setup

For the purpose of comparison and evaluation, we also build a second prototype using an off-the-shelf camera (Panasonic DMC-FZ72) and a light-box, as shown in Figure 4. A piece of glass is covered on top of the banknote to keep it flat. A light-box brightens up the banknote from the underneath so that the camera can photograph the translucent patterns on the top at a close distance (about 2 cm). The aperture of the camera is fixed at 5.0 and the shutter speed at 1/100. This combination provides sufficient depth of field as well as stability to get a clear and sharp image. The shooting mode is set to "Macro" to capture the details of random patterns in a close-up.

### C. Image Processing

After we photograph a back-lit polymer banknote, the image is cropped to contain only the feature area, which is located by the aide of auxiliary markers. The cropped image is further processed by applying 2-D Gabor filters into a compact 2048-bit binary code, which we call a *polymer substrate fingerprint*. Details of this process are explained below.

*1) Gabor Filter Selection:* Two-dimensional Gabor filters are a common technique used to analyse the textural patterns of an image. They have been commonly employed in biometric applications such as iris and face [10]. A 2-D Gabor filter comprises a sinusoidal wave modulated by a Gaussian envelope. It efficiently detects the edges and textural patterns existing in a 2-D image by capturing features in both frequency and spatial domains. This allows the output of the 2-D Gabor filter to be used in distinguishing whether the two snapshots are originally from the same pattern. In our work, we only need it to work in the spatial domain. In this domain, a 2-D Gabor filter is described as below [10]:

$$\begin{aligned} \psi(x,y) &= \frac{F^2}{\pi\gamma\eta} e^{-F^2\left[(x'/\gamma)^2+(y'/\eta)^2\right]} e^{i2\pi F x'} \\ with: & \\ x' &= x\cos(\theta) + y\sin(\theta) \\ y' &= -x\sin(\theta) + y\cos(\theta), \end{aligned} \tag{1}$$

where $F$ is the central frequency of the sinusoidal wave, $\theta$ is the angle between the direction of the wave and the $x$ axis of the spatial domain, $e$ is the natural exponential function, $\gamma$ and $\eta$ are the standard deviations of the Gaussian envelope in the direction of the wave and orthogonal to it, respectively. The parameters $\gamma$ and $\eta$ represent the shape factors of the Gaussian surface, and are also called the *smoothing* parameters. They determine the selectivity of the filter in the spatial domain.

Different combinations of the Gabor filter parameters are capable to extract different textural features. However, there is no unified way to determine values for these parameters [10], as they depend on particular characteristics of the textural patterns to be extracted [11]. To efficiently select the combination, a matrix called a Gabor filter-bank is created that contains a range of frequencies and orientations of Gabor filters. Each individual frequency in the matrix is called a *scale*, which is calculated from a maximum frequency, known as $f_{\max}$. For a total of $U$ frequencies, each scale is defined as follows:

$$scale = \frac{f_{\max}}{\sqrt{2}^{u-1}}, \quad \forall u \in \{1, 2, \ldots, U\}. \tag{2}$$

For a total number of $V$ orientations, each orientation is calculated as follows:

$$orientation = \frac{v-1}{V}\pi, \quad \forall v \in \{1, 2, \ldots, V\}. \tag{3}$$

Suitable parameters for the Gabor filters can be determined by using an iterated process through experiments [10]. Once a suitable set of parameters is found, it can be used for the same type of textural patterns (e.g., using the same set of parameters for processing all human irises in iris recognition).

When choosing the values for the Gabor filter parameters, we have three main considerations. First of all, we consider the *decidability* [12], which measures how far the clustering of samples from the same source is statistically separated from the clustering of samples from different sources. Clearly, the decidability should be sufficiently large. Second, we consider the fractional Hamming distance (HD), which represents the percentage of bits that are different on corresponding bit positions between two binary strings. In the rest of the paper, we will use HD as a shorthand to refer to fractional Hamming distance. The HD between samples from different polymer notes should ideally centre around 0.5. As we will show in the evaluation, centring around 0.5 will greatly simplify our analysis as the obtained binary fingerprint can be modelled as a series of Bernoulli trials. Third, after the image processing, the obtained polymer fingerprint should contain sufficiently high entropy. A high entropy (say more than 128 bits) will statically guarantee that the chance for a random polymer substrate to successfully pass the verification is negligible. In fact, as we will demonstrate, we are able to achieve much higher entropy (900 bits) in the extracted fingerprints. Based on these requirements and the selection method outlined in [10], we conduct empirical experiments based on 100 samples from a set of randomly chosen £10 banknotes and determine that a suitable set of parameters for extracting the random translucent patterns for a polymer substrate is $f_{\max} = 0.25$, $\gamma = \sqrt{2}$, $\eta = \sqrt{2}$. The values for the scale and orientation that give the best overall performance are $u = 5$ and $U = 6$ for computing the scale (Equation 2) and $v = 11$ and $V = 30$ for computing the orientation (Equation 3). This setting is the same for both £5 and £10 as the textural patterns are the same. For paper notes, we use a different combination: $u = 6$ and $U = 6$ for the scale, and $v = 22$ and $V = 25$ for the orientation. The parameters are slightly different because of the different textural patterns exhibited by a paper note (see Figure 2 (d)).

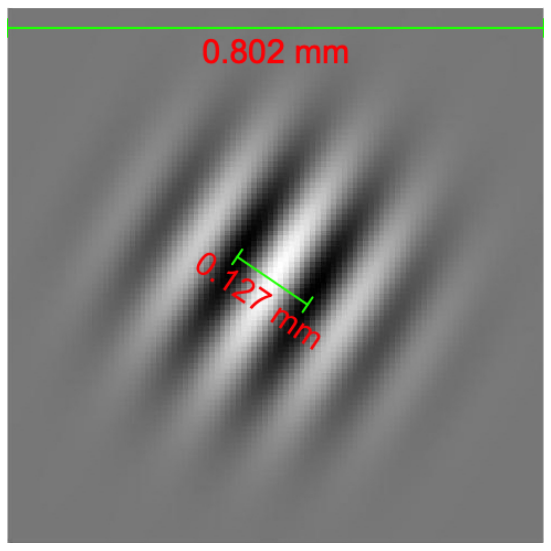

Fig. 5: Physical dimension of the Gabor filter

The size of the Gabor filter applied on the scanned polymer banknote is $101 \times 101$ (unit: pixel). Given the resolution of the scanner being 3200 dpi, each pixel in the scanned sample corresponds to about 7.94 $\mu$m (1/3200 inch). For 101 pixels, that corresponds to a physical size of $101 \times 7.94 = 802$ $\mu$m on the banknote. According to Equation 2, the frequency of the Gaussian envelope applied on the polymer banknote is a quarter of $f_{\max}$. Therefore, the wavelength is 16 pixels, equating to 127 $\mu$m as shown in Figure 5.

*2) Feature Extraction and Comparison:* With the 2-D Gabor filter defined above, we apply it to process the translucent patterns photographed from the feature area of a polymer note into a binary string of 2048 bits, similar to how an iris-code is generated from the textural patterns of an iris image in iris recognition [12]. First of all, a captured photograph is grey-scaled, and a 2-D Gabor filter kernel is applied on the converted image to obtain a matrix of complex numbers. Each pixel in the image is transformed into a complex number. Given an input image $I(x, y)$ of dimensions $X \times Y$ and a bank of discrete Gabor filters $G_{mn}(x, y)$ with $m \in \{1, .., M\}$ and $n \in \{1, .., N\}$, the complex number matrix $C(x, y)$ is computed for each filter of the bank as follows:

$$C_{mn}(x, y) = \sum_{a=1}^{X} \sum_{b=1}^{Y} I(a, b) \overline{G}_{mn}(x - a, y - b), \quad (4)$$

where $\bar{}$ denotes the complex conjugate.

Because the values of adjacent pixels are usually highly correlated, we perform a down-sampling process in order to remove the correlation. Values in every $20^{\text{th}}$ rows and $20^{\text{th}}$ columns are selected to form a new matrix sized $32 \times 32$. All elements in the matrix are complex numbers with real and imaginary parts. Each element is then decoded into 2 bits depending on which quadrant does the complex number falls into. This gives a binary output of $32 \times 32 \times 2 = 2048$ bits, which we call a “polymer substrate fingerprint”.

The similarity between two polymer substrate fingerprints, denoted as $f_1$ and $f_2$, is measured by computing a fractional Hamming distance based on an XOR $\oplus$ operation, as below:

$$HD = \frac{|f_1 \oplus f_2|}{2048} \quad (5)$$

This is similar to how iris-codes are compared in iris recognition [12], however, in the case of the iris, there is a 2048-bit mask vector in addition to a 2048-bit iris-code. The purpose of the mask is to filter out unreliable bit positions caused by artefacts such as eyelids and eyelashes from the HD computation. In our system, we carefully select a feature area that is not obstructed or interfered by artefacts such as holograms and other printed security features. This removes the need for a mask. Hence, the stored data is only half the size of an iris-code. In the ideal case, the HD between any two fingerprints extracted from the same banknote should be close to 0, and the HD between fingerprints extracted from different banknotes should be close to 0.5. In the sections below, we will systematically evaluate the HD comparison results.

TABLE I: Summary of Datasets

| Group | Condition | Equipment | Denom. | Num. |
|---|---|---|---|---|
| Benchmark | Favourable | Scanner | £10 | 1000 |
| Robustness Test | Rotation | | | |
| | Scribbling | | | |
| | Soaking | | | 100 |
| | Folding | | | |
| Variation Test | Equipment | Camera Set | | 1000 |
| | Denomination | Scanner | £5 | |
| | Substrate | | £20 | |

## IV. Datasets

We collect an extensive set of samples from the UK banknotes of different denominations, under different conditions. In total, we have collected 8 datasets containing 6,200 sample images, taken from 340 different banknotes, including 140 £10 notes, 100 £5 notes and 100 £20 notes. These datasets can be divided into three groups: benchmark, robustness test and variation test, as summarised in Table I.

### *A. Benchmark*

The dataset in the benchmark group is collected in a favourable condition. It consists of 100 different £10 polymer notes with 10 image samples for each note, making it a total of 1,000 samples. Each sample banknote is sandwiched between two pieces of thin clear glasses in the aligned frame during the scanning process. The use of the frame helps constrain the banknote in the correct orientation.

## *B. Robustness Test*

**Rotation**. As part of the robustness test, we rotate the banknote and use the auxiliary markers to automatically re-orient the image before processing. This is done in Matlab. We collect 10 samples per each banknote from the same £10 banknotes used in the benchmark set, after rotating each note by a different angle varying from $-10°$ to $10°$. Testing rotation within this range is sufficient for our purpose as in practice errors of mismatch occur by only a small rotation angle. In the two dimensional space, given coordinates of two points $(x_1, y_1)$ and $(x_2, y_2)$ in a Cartesian coordinate system, the orientation angle $\alpha$ is calculated below:

$$\alpha = \tan^{-1} \frac{|x_1 - x_2|}{|y_1 - y_2|} \tag{6}$$

The angle of the rotation $\alpha$ for each sample image is computed based on the centres of the two auxiliary markers. Then the image is rotated accordingly.

**Scribbling**. Under the Currency and Banknotes Act 1928 in the UK, it is prohibited to scribble on the surface of banknotes as that may deface the notes. Therefore, we use hairs and fibres attached to the surface of each £10 banknote used in the benchmark dataset to mimic the same effect of scribbling when the banknote is photographed.

**Soaking**. Sometimes a banknote may drop into water by accident, or get wet (e.g, by rain) during the daily usage. Because every polymer banknote is protected by a over-coating layer (varnish) as part of the production process, a polymer note is water-resistant by design. Nonetheless, we use twenty randomly selected £10 to conduct a soaking test, with one sample for each note taken before the test and four samples taken after the test. These banknotes are soaked in water for 2 minutes and then dried naturally on a flat surface for 30 minutes before they are scanned and processed.

**Folding**. In daily life, banknotes are often folded before being put in a wallet. We conduct a test to study the effect of folding on our method. Initially, we take a set of randomly selected £10 notes, fold each note in half and store them in a daily used wallet for three days. Afterwards, the folded notes are flattened with the images of the feature area taken. Next, we fold each banknote twice along the long side to make it more compact for storage in the wallet. The double folded banknotes are put in the wallet for another three days before they are flattened and scanned. The folding dataset consists of 100 sample images taken from twenty £10 with one sample of the original note, two samples after folding once, and another two samples after folding twice.

## *C. Variation Test*

**Alternative Equipment**. Instead of a film scanner, we use a camera and a light-box to photograph the same 100 £10 notes used in the benchmark set with 10 images for each note. Film scanners and cameras are two different types of optical imaging devices, using different physical mechanisms. A film scanner obtains an image by moving a bar of light sensors alongside the surface of a flat film with a light shining on the opposite side of the film, while a camera flashes an array of light sensors in one go. Despite having a slow developing speed, the scanner tends to capture a high-quality edge-to-edge image. The reason is that it has a relatively simple optical structure with only one flat protective screen being laid on top of the sensor, while for a camera, light needs to pass through 4 to 7 lenses before reaching the sensors. The polymer fingerprints obtained from using these two different devices will be compared in the valuation section.

**Different Denominations**. The £5 and £10 banknotes use essentially the same polymer substrate. Under the microscopic view, we observe similar random translucent patterns in the opacity coating layer for both £5 and £10 notes. To study the variation between these notes of different denominations, we use the film scanner to photograph 100 £5 polymer notes with 10 samples per note, and compare them against the benchmark set. We use the same Gabor filter setting for £5 as used for £10 in the benchmark dataset.

**Different Substrates**. To study of the variation between a polymer substrate and a paper substrate, we randomly choose 100 £20 paper notes. The paper £20 note in the UK uses a paper substrate made of cotton and linen. We use the same film scanner to image 100 £20 notes with 10 samples per banknote. As we will show in the evaluation, although our technique is designed for the anti-counterfeiting of polymer notes, it can be easily adapted to prevent forgery of paper notes as well.

# V. Evaluation

## *A. Framework*

Our polymer substrate fingerprinting technique is closely related to the technology of biometrics which authenticates people based on their inherent physical or behavioural features. Here, we authenticate a polymer banknote based on its inherent physical properties in the polymer substrate. On the other hand, our method is also related to the field of physically unclonable function (PUF), which provides security assurance based on the impossibility to physically clone a physical object. However, biometrics and PUF generally use different evaluation metrics despite that the two are inherently related. Based on earlier work [11], we propose to use a unified framework that combines both biometrics and PUFs metrics for evaluating our polymer substrate fingerprinting system.

*1) Biometrics:* A biometric system authenticates people based on their unique physical or behavioural features [12]. The performance of a biometric, especially one that uses HD for comparison, is commonly evaluated in terms of decidability, degree of freedom, and error rates as explained below.

**Decidability.** In a biometric system, there are two groups of biometric data distributions: the intra-group that refers to the distances between samples from the same subject and the inter-group that refers to the distances between samples from different subjects. In this paper, we use fractional Hamming distance (HD) as an example of the distance metric. Clearly the two distributions should be as further apart as possible. We use the *decidability* metric [12] to measures how far the two distributions are separated. This metric is denoted $d'$ and is computed as below:

$$d' = \frac{|\mu_1 - \mu_2|}{\sqrt{\frac{\sigma_1^2 + \sigma_2^2}{2}}}, \tag{7}$$

where $\sigma_1$ and $\sigma_2$ are the standard deviations of distances between samples from the intra-group and the inter-group, respectively, $\mu_1$ and $\mu_2$ are the mean values from these two groups. $|\cdot|$ denotes the absolute value.

TABLE II: Notations used in PUF metrics. (In the benchmark dataset, $S$ = 100, $L$ = 2048 and $T$ = 10)

| | |
|---|---|
| $f$ | Feature vector |
| $S$ | Total number of banknotes |
| $s$ | Index of each banknote ($1 \leq s \leq S$) |
| $L$ | Bit Length of the feature vector from each banknote |
| $l$ | Index of each bit position in a feature vector ($1 \leq l \leq L$) |
| $T$ | Total number of samples measured per banknote |
| $t$ | Index of each sample ($1 \leq t \leq T$) |

**Degree of Freedom.** The number of degrees of freedom (*DoF*) is a metric that measures how many independent bits exist in a biometric instance. In our systems, the more degrees of freedom contained in the extracted feature vectors, the more statistically unlikely it will be for any two random feature vectors to match. The *DoF* is calculated below [12]:

$$N = \frac{\mu(1-\mu)}{\sigma^2}, \tag{8}$$

where $\mu$ is the mean of the *HD* in the inter-group, and $\sigma$ is the standard deviation of the *HD* in this group.

**Error Rates.** In a biometric verification system, there are two types of error rates: a false rejection rate (*FRR*) and a false acceptance rate (*FAR*). *FRR* refers to the probability that a genuine sample is falsely rejected, while *FAR* refers to the probability that a fake sample is falsely accepted. For practical purposes, both *FRR* and *FAR* should be kept as small as possible (ideally 0%). In reality, they vary according to the choice of a threshold. Increasing the threshold can reduce *FRR* but often at the expense of increasing *FAR*. Commonly an equal error rate (EER), where the curves of *FRR* and *FAR* intersect, is used to indicate the overall error rate performance of a biometric system.

*2) Physical Unclonable Function:* Physical Unclonable Function (*PUF*) is a security primitive built upon the difficulty of replicating the same physical properties of an object or device. Maiti et al. [13] proposed a framework to evaluate the performance of PUF. We adapt their framework as part of the metrics used to evaluate the performance of our system in the following three dimensions: space, time, and device. Notations used in this framework are summarised in Table II.

**Space Dimension - Uniformity, Randomness.** In the space dimension, we assess how uniform the 0s and 1s are distributed in a feature vector and how random the binary values are at each bit position of a feature vector.

$$Uniformity(s,t) = \frac{1}{L}\sum_{l=1}^{L} f_{s,t,l} \tag{9}$$

$$\begin{aligned} Randomness(s) &= -\log_2 \max(p_s, 1-p_s) \\ where \quad p_s &= \frac{1}{TL}\sum_{t=1}^{T}\sum_{l=1}^{L} f_{s,t,l} \end{aligned} \tag{10}$$

**Time Dimension – Reliability, Steadiness.** In the time dimension, we assess the similarity of samples taken at different times from the same banknote. Reliability measures how consistent a feature vector from a banknote is as compared with other feature vectors taken in different times from the same banknote. Steadiness measures how stable the value at each bit position is among all feature vectors taken from the same banknotes.

$$Reliability(s) = 1 - \frac{2}{T(T-1)L}\sum_{t=1}^{T-1}\sum_{t'=t+1}^{T}\sum_{l=1}^{L}\left(f_{s,t,l} \oplus f_{s,t',l}\right) \tag{11}$$

$$\begin{aligned} Steadiness(s) &= 1 + \frac{1}{L}\sum_{l=1}^{L}\log_2 \max\left(p_{s,l}, 1-p_{s,l}\right) \\ where \quad p_{s,l} &= \frac{1}{T}\sum_{t=1}^{T} f_{s,t,l} \end{aligned} \tag{12}$$

**Device Dimension – Uniqueness, Bit-Aliasing.** In the device dimension, we consider the diversity of the feature vectors taken from different banknotes. Uniqueness measures how distinguishable a feature vector is from other feature vectors extracted from different banknotes. Bit-aliasing measures how likely different banknotes are to produce identical values at the same bit positions in the feature vector.

$$Uniqueness(s) = \frac{2}{T^2 S(S-1)L} \cdot \sum_{t=1}^{T}\sum_{\substack{s'=1 \\ s'\neq s}}^{S}\sum_{t'=1}^{T}\sum_{l=1}^{L}\left(f_{s,t,l} \oplus f_{s',t',l}\right) \tag{13}$$

$$Bit\text{-}Aliasing(l) = \frac{1}{ST}\sum_{s=1}^{S}\sum_{t=1}^{T} f_{s,t,l} \tag{14}$$

### B. Results

*1) Benchmark Performance:* Based on the benchmark dataset, we compute pair-wise HDs between the feature vectors obtained from the same banknotes (intra-group) and from different banknotes (inter-group). The histograms for the two groups of HD calculations are plotted in Figure 6.

**Biometric Metrics.** From Figure 6, the inter-group and intra-group HD distributions are clearly separated. Based on Equation 7, we calculate the *decidability* $d' = 29$, which is much larger than the reported $d' = 14$ from iris codes [12]. One main reason for the higher decidability in our system is that we photograph the random features of a polymer substrate at an extremely close distance (1-2 cm), but this is not possible with the iris scanner as that would be too invasive to a human.

For the inter-group HD distributions, we obtain the mean HD $\mu = 0.500$ with a standard deviation $\sigma = 0.017$. Based on Equation 8, we are able to calculate the number of degrees of freedom $N = 900$. To confirm that $N$ accurately reflects the number of degrees of freedom for the actual polymer fingerprints, we plot a bionomial distribution curve which models a series of 900 Bernoulli trials (i.e., tossing an unbiased coin) with a probability of 0.5 for each trial. As shown in Figure 6, this binomial distribution curve fits perfectly the HD histogram in the inter-group. This corroborates the fact that the obtained 2048-bit fingerprints from the polymer banknotes have 900 degrees of freedom, or in other words 900 bits entropy. By comparison, the number of the degrees of freedom for a 2048-bit iris code is only 249 [12]. Note that the iris textural patterns tend to be correlated along the radial directions [12], which reduces the entropy of the iris codes, while such correlations do not exist in the polymer substrate. This, together with the fact that we can take a close-up of the polymer substrate at an extremely short distance, contributes to the much higher entropy in the extracted polymer substrate fingerprints than in iris-codes. The intra-group HD distributions do not show the same symmetric shape as the inter-group HD distributions as they heavily depend on the noise in the data acquisition. A few noisy samples can result in relatively high intra-group HDs for the same banknotes, leaving a long trail in the distribution.

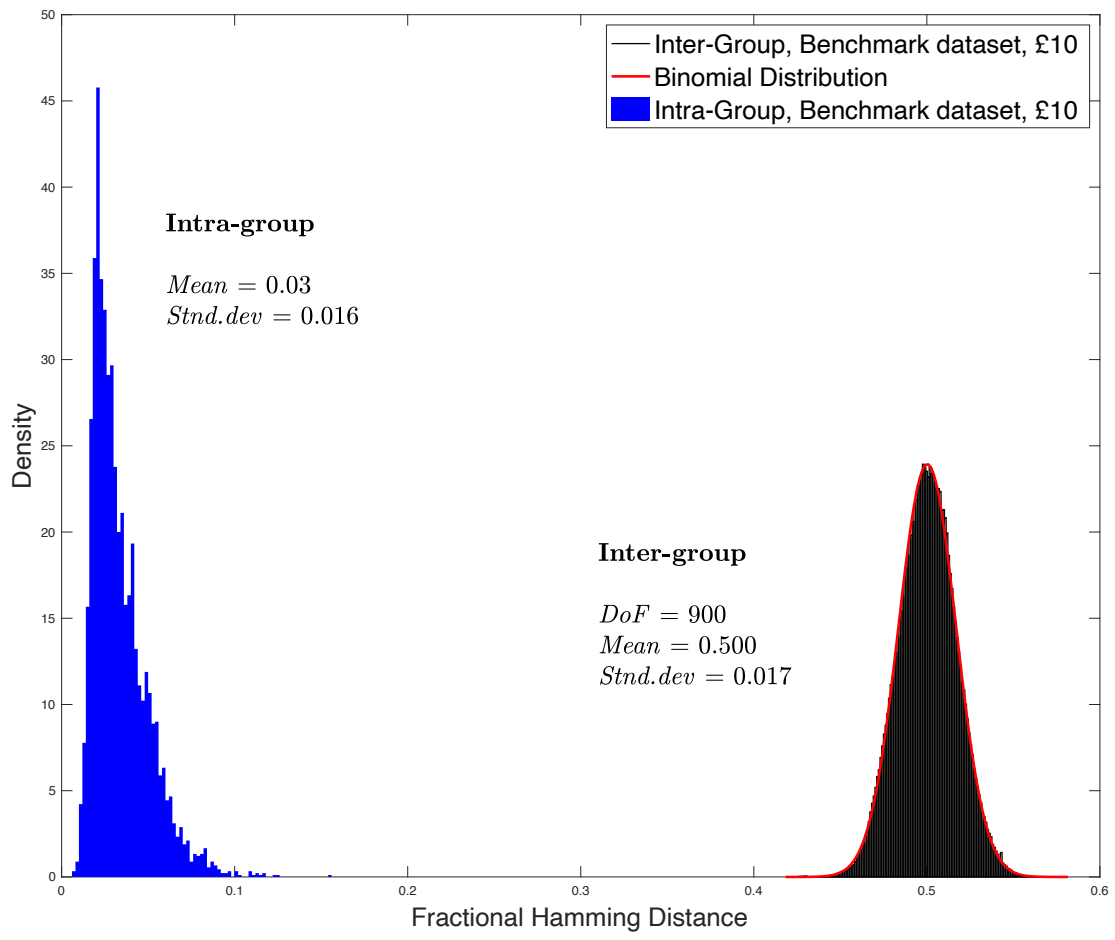

Fig. 6: HD distributions of Benchmark dataset. Decidability $d' \approx 29$.

From Figure 6, it is clear that the two groups of distributions are far apart. If we choose an HD value 0.33 as the threshold, the *FRR* and *FAR* will be both 0%. Obviously the *EER* for the overall performance is also fixed at an ideal value 0%.

**PUF Metrics.** In Section V-A2, we have defined a set of metrics to evaluate PUF. Table III summarises the performance of polymer substrate fingerprints using those metrics due to Maiti et al. [13] along with other related PUFs proposed in the past work for comparison. As shown in Table III, our technique achieves results close to the ideal values in each of these metrics. Overall the results also compare favourably in general to the state-of-the-art PUF systems reported in the literature [11], [13].

*2) Robustness Tests:* For ease of illustration, we plot HD histograms for different robustness test cases as fitted curves in Figure 7. We explain each case below.

**Rotated Dataset.** As shown in Figure 7, *rotation* has little effect on the performance as the software is able to automatically re-orient a banknote image based on auxiliary markers before the feature area is processed. Both the intra-group and inter-group distributions remain largely unchanged. As an example, if we choose HD = 0.33 as the threshold, the *FRR* and *FAR* still maintain at 0%.

TABLE III: PUF metrics from Benchmark dataset

| PUF Metrics | Ideal Value | Bench. Dataset | Paper PUF [11] | Arbiter PUF [13] | Ring Oscilator PUF [13] |
|---|---|---|---|---|---|
| Uniformity | 0.5 | 0.500 | 0.466 | 0.556 | 0.505 |
| Randomness | 1 | 0.980 | 0.907 | 0.846 | 0.968 |
| Steadiness | 1 | 0.962 | 0.945 | 0.984 | 0.985 |
| Reliability | 1 | 0.967 | 0.938 | 0.997 | 0.991 |
| Uniqueness | 0.5 | 0.500 | 0.465 | 0.072 | 0.472 |
| bit-Aliasing | 0.5 | 0.500 | 0.466 | 0.195 | 0.505 |

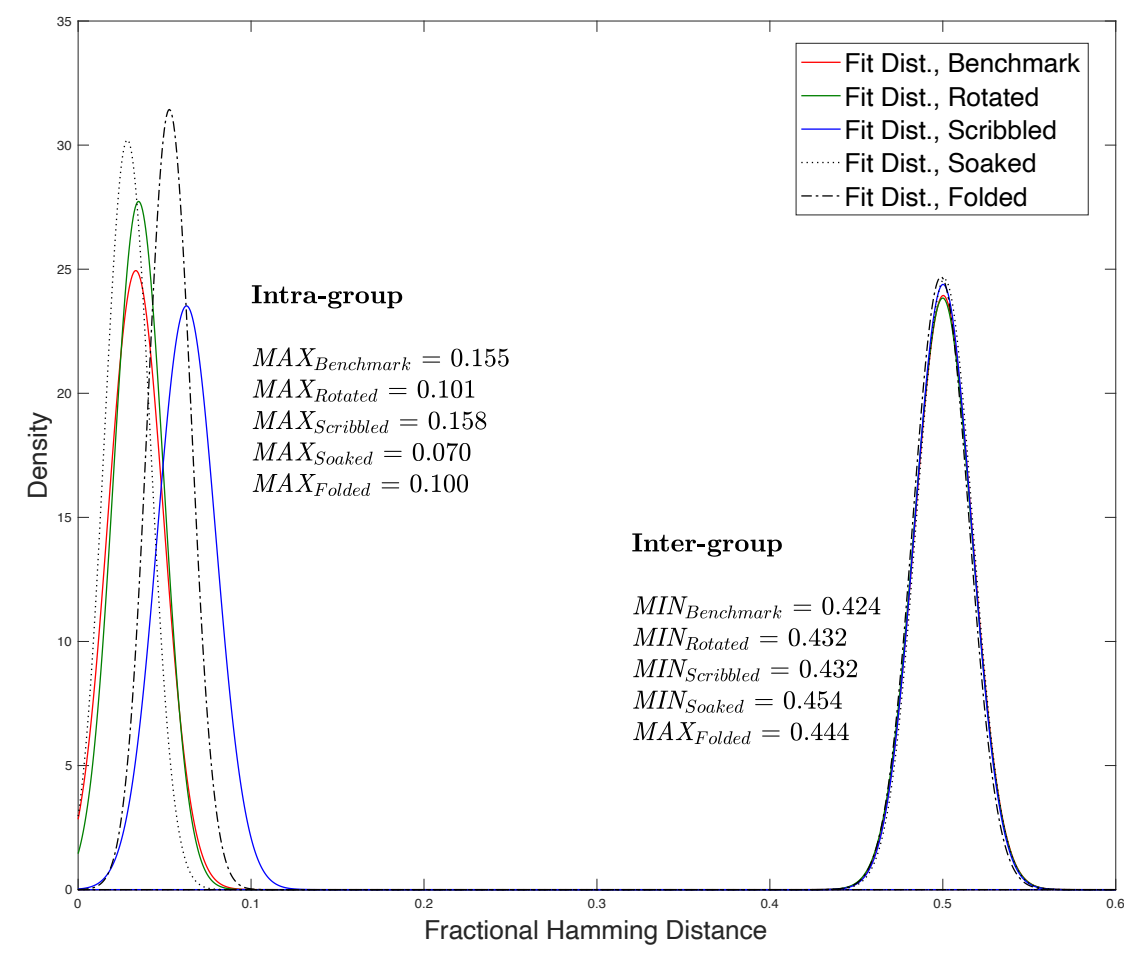

Fig. 7: HD histograms after robustness tests with reference to benchmark

**Scribbled Dataset.** As compared to the benchmark dataset, scribbling on the banknotes shifts the centre of the intra-group distribution to the right (from 0.03 to 0.06), but it has little effect on the inter-group distribution. This means scribbling on a banknote adds noise to the data, but the two groups remain clearly separated. At a threshold of HD = 0.33, the *FRR* and *FAR* are kept at 0%.

**Soaked Dataset.** As shown in Figure 7, soaking a banknote has little effect on both the intra-group and inter-group distributions. This is as expected since the polymer banknotes are water-proof by design (due to the application of clear veneer at the outer layer). Given HD = 0.33 as the threshold, the *FRR* and *FAR* are still 0%.

**PUF Result.** As shown in Table IV, the robustness tests in our experiments have little effect on the PUF metrics. All the values computed after the robustness tests remain close to the ideal values. This suggests that our technique is reasonably robust against non-ideal daily handling of banknotes. To a large extent, the strong robustness of our method is attributed to the basic design of a polymer note: in particular, the veneer coating at the outer layer protects the printing underneath and makes the polymer note highly durable against rough daily usage.

*3) Variation Tests:* In the section, we study the variation of performance under different test conditions, including the use

TABLE IV: PUF metrics after robustness tests vs benchmark

| PUF Metrics | Ideal Value | Rotated Dataset | Scribbled Dataset | Soaked Dataset | Folded Dataset | Bench. Dataset |
|---|---|---|---|---|---|---|
| Uniformity | 0.5 | 0.499 | 0.499 | 0.498 | 0.501 | 0.500 |
| Randomness | 1 | 0.981 | 0.980 | 0.978 | 0.978 | 0.980 |
| Steadiness | 1 | 0.960 | 0.960 | 0.971 | 0.949 | 0.962 |
| Reliability | 1 | 0.965 | 0.965 | 0.972 | 0.950 | 0.967 |
| Uniqueness | 0.5 | 0.500 | 0.500 | 0.501 | 0.501 | 0.500 |
| bit-Aliasing | 0.5 | 0.499 | 0.499 | 0.498 | 0.501 | 0.500 |

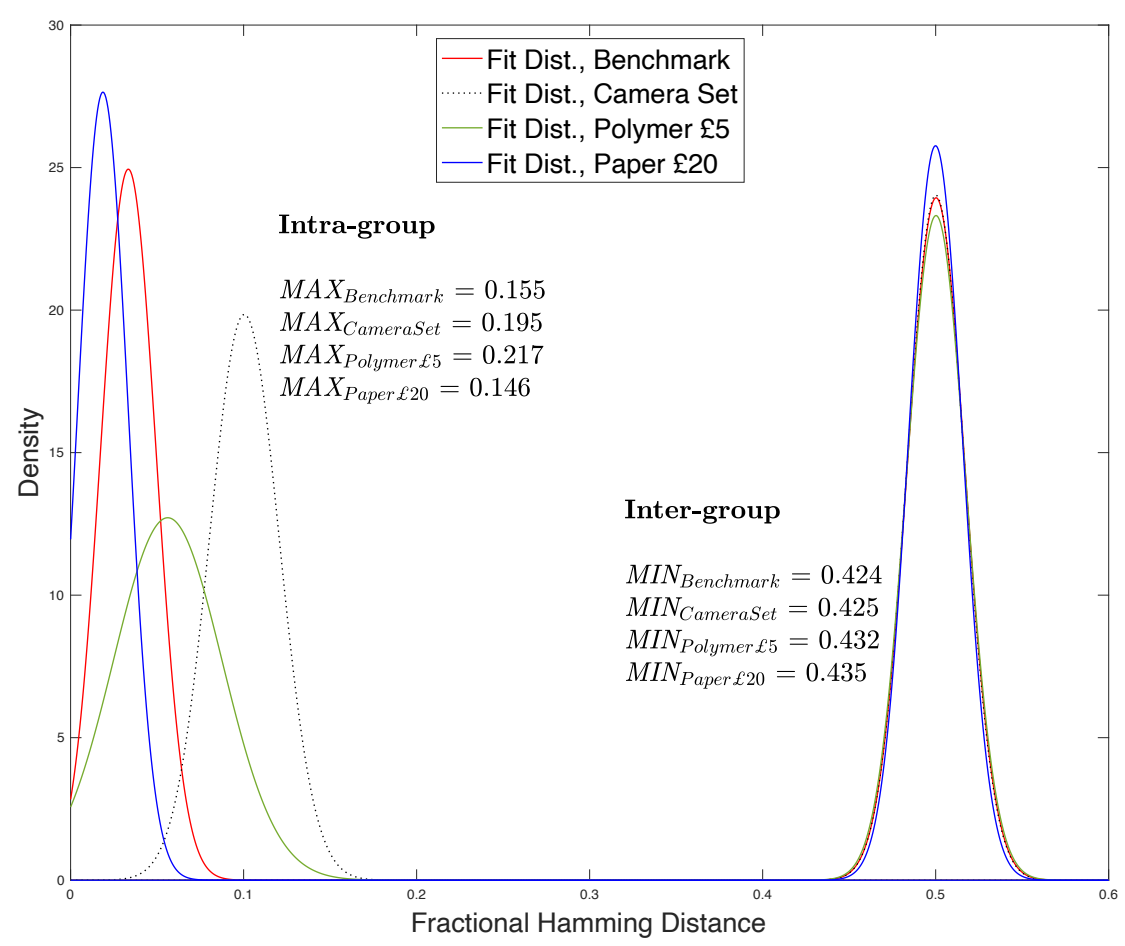

Fig. 8: Histograms of variation tests vs benchmark

TABLE V: PUF metrics after variation tests vs benchmark

| PUF Metrics | Ideal Value | Camera Set | Polymer £5 | Paper £20 | Benchmark Dataset |
|---|---|---|---|---|---|
| Uniformity | 0.5 | 0.499 | 0.499 | 0.497 | 0.500 |
| Randomness | 1 | 0.984 | 0.985 | 0.983 | 0.980 |
| Steadiness | 1 | 0.884 | 0.935 | 0.978 | 0.962 |
| Reliability | 1 | 0.900 | 0.944 | 0.981 | 0.967 |
| Uniqueness | 0.5 | 0.500 | 0.500 | 0.500 | 0.500 |
| bit-Aliasing | 0.5 | 0.499 | 0.499 | 0.497 | 0.500 |

of a different imaging device, a polymer note of a different denomination and a banknote of a different substrate. Fitted curves for the HD histograms under these different conditions are shown in Figure 8. Values of the PUF metrics calculated under these conditions are summarised in Table V.

**Alternative Equipment.** When a camera and a light-box are used instead of a film scanner to photograph the feature area, the steadiness and the reliability of the obtained feature vector slightly decrease as shown in Table V. This is also reflected in Figure 8, in which the intra-group distribution slightly shifts to the right. As a result, the decidability $d'$ is reduced from 30 in the benchmark set to 22. This is because the camera used in the experiment has a more complex optical path for light reaching the *CMOS* sensors than the film scanner. Furthermore, a close-up taken by a camera in the macro mode under a close distance (about 2 cm) tends to be slightly bent near the edge of the ring [11], which adds noise to the feature extraction. Nonetheless, the distributions of the two groups are still clearly separated with a maximum *HD* of 0.195 for the intra-group, and a minimum *HD* of 0.425 for the inter-group. The *FRR* and *FAR* remain at 0% when the threshold is set to 0.33.

**Different Denominations.** As compared to £10, the intra-group HD distributions for £5 shifts to the right as shown in Figure 8, while the inter-group distribution remains basically unchanged. This is mainly because the £5 polymer note is physically smaller than the £10 polymer note, and the area suitable for feature extraction (i.e., directly exposed from the opacity coating layer without the obstruction of security printing) is also smaller. In our experiment, while the Gabor filter setting is the same, the feature area defined for £5 is only about half of the area for £10 (also see Figure 2). While the smaller area has little impact on the PUF metrics (see Table V), it reduces the decidability $d'$ from 30 to 18, and the *DoF* from 900 to 854.

**Different Substrates.** Since paper £20 notes are still used in the UK, we test our fingerprinting technique on £20 notes that use paper substrate. We obtain slightly better performance than the benchmark £10 polymer notes. The decidability $d'$ is slightly increased from 30 to 32, while the *DoF* is increased from 900 to 1043. The *FRR* and *FAR* remain at 0% for the threshold of HD = 0.33.

**PUF Result.** The PUF metric values are basically the same as the benchmark dataset. The slightly better performance of £20 is related to its inherent textural patterns. As shown in Figure 2, a paper £20 banknote also exhibits random translucent patterns, which are caused by the random leaving of the cotton fibre and linen rather than the opacity coating, but the image seems to contain richer textural information than a polymer substrate. This shows that although our technique is designed for the anti-counterfeiting of polymer notes, it can also be adapted to prevent forgery of traditional paper notes.

### C. Limitations

Our current work has a few limitations. First all, the data samples are taken from the UK banknotes only. Given that the manufacturing of polymer notes follows essentially the same process, we believe the results are applicable to banknotes in other countries, but this needs to be confirmed in further research. Second, we have done robustness tests under common cases, but the tests are not exhaustive. Further evaluation may include folding the banknote more than twice, placing the banknote under high temperature (near the melting point), and studying the effect of wearing out after years of usage. Finally, the features are extracted from different areas on banknotes of different denominations. Hence, the system needs to identify the denomination first, which is doable but adds an extra step in the processing. Defining a standardised feature area for all polymer banknotes will be highly desirable.

## VI. Anti-counterfeiting Applications

### A. Online Application

First of all, we propose an online application, which works with an existing *unmodified* banknote. Here, we will leverage the fact that each banknote has a unique serial number, as we will explain below.

We divide an online application into two phrases: registration and verification. During the registration phase, a polymer substrate fingerprint for each newly manufactured polymer banknote is extracted and recorded in a database along with a unique serial number of the banknote. In the verification phase, a fresh photograph of the feature area is taken and processed into a compact 2048-bit fingerprint. The fingerprint, along with the banknote serial number, is then sent to a remote server through a secure channel (e.g., SSL/TLS). Based on the serial number, the server retrieves the reference fingerprint and compares it with the sample fingerprint against a HD threshold. Finally, the verification result is communicated back to the client through the existing secure channel.

TABLE VI: False match for one-to-one comparison

| HD threshold | Odds of False Match |
|---|---|
| 0.3 | $3.5 \times 10^{-34}$ |
| 0.31 | $6.0 \times 10^{-31}$ |
| 0.32 | $6.7 \times 10^{-28}$ |
| 0.33 | $5.0 \times 10^{-25}$ |
| 0.34 | $2.5 \times 10^{-22}$ |
| 0.35 | $8.2 \times 10^{-20}$ |
| 0.36 | $1.9 \times 10^{-17}$ |
| 0.37 | $2.9 \times 10^{-15}$ |
| 0.38 | $3.0 \times 10^{-13}$ |
| 0.39 | $2.2 \times 10^{-11}$ |
| 0.4 | $1.1 \times 10^{-9}$ |

Thanks to the unique serial number, the verification in the online application is based on one-to-one comparison (rather than one-to-many as required in exhaustive search). This is not only extremely fast, but also gives great flexibility in choosing a threshold. The false rejection rate of a system heavily depends on the data acquisition environment during the verification. On the other hand, the false acceptance rate is essentially determined by the inherent entropy of the data source, and we can theoretically estimate the value as follows. Let $P_a$ be the false acceptance rate of a fingerprint for one-to-one comparison. Based on the 900 degrees of freedom and the binomial distribution fitting in Figure 6, we model each fingerprint as the result of performing a series of $N = 900$ Bernoulli trials with the probability $p = 0.5$ of guessing 'heads' (or 'tails') correctly for each trial. Hence, we can compute $P_a = \sum_{i=0}^{m} N!/(m!(N-m)!) \cdot p^m \cdot (1-p)^{N-m}$, where $m$ is the number of successful guesses [12]. Given a threshold $\theta = 0.33$, $m \approx \theta \cdot N$, the results are summarised in Table VI. As shown in the table, even if we set the threshold to be HD = 0.4 to give more tolerance to intra-group variations, the false acceptance rate remains negligible.

### B. *Offline Application*

An offline application differs from an online one by printing the registration information onto a banknote rather than saving it to a database. However, this adds an extra step of registration to the existing banknote manufacturing process. Figure 9 summarises the process of the registration. The feature vector extracted from the translucent patterns of a polymer substrate is digitally signed, along with other contextual information such as the serial number and denomination value. The private signing key is kept securely by the authorities who issue banknotes. In the proof-of-concept implementation, we use ECDSA with 512 bits key length (256-bit security) for digital signing. Encoded in Base64, the total length of the message and the digital signature is approximately 4420 bits, which can be fit into a QR code (version 18) with medium error correction (as shown in Figure 9).

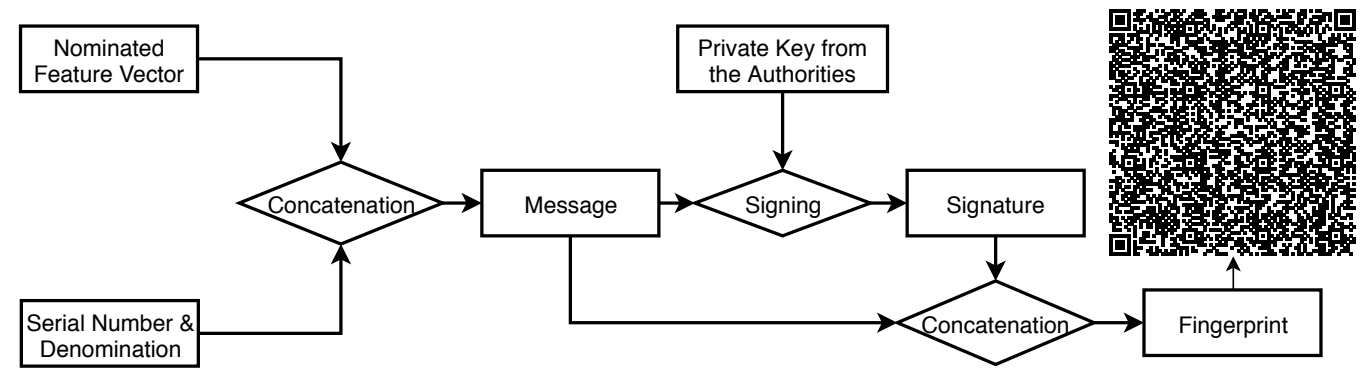

Fig. 9: The Procedure of Fingerprint Registration. The QR code shown in the diagram is generated from a real £10 polymer banknote.

Fuzzy encryption is a common technique to encrypt a biometric sample such that it can only be decrypted by another biometric sample taken from the same subject [14]. We could apply the same technique to encrypt the fingerprint contained in the QR code, however, in the context of our application, a counterfeiter always has physical access to real banknotes which they wish to counterfeit. Hence, encrypting the fingerprint does not offer real security benefit in our case. For this reason, we simply save the fingerprint in its plain text in the QR code, but we add a digital signature to protect the integrity of data.

During the verification phase, data from the QR code is first read, which contain a reference fingerprint, a digital signature and other data. After the digital signature is verified successfully by using a public key, a fresh image of the feature area is taken and then processed into a sample fingerprint. This fingerprint will be compared with the reference fingerprint against a HD threshold with a binary outcome: *accept* or *reject*. The integrity of the content in the QR code is protected through the digital signature. If an attacker copies the same QR code to a different banknote, the verification will fail as the two fingerprints will not match.

## VII. Security Analysis

### A. *Threat Model*

We assume that the attacker knows everything about the fingerprinting of banknotes. He knows the full state of the device and exactly which area is used for feature extraction and how the feature vector is computed from the feature area. We further assume that the attacker has access to effectively the same printing material and equipment as used for producing legitimate banknotes. Under this assumption, the security of the existing polymer banknotes will be easily broken. While the security of existing banknotes is easily broken in this threat model, we aim to provide additional security assurance such that counterfeiting remains difficult.

The attacker has several limitations. First of all, we assume he is unable to obtain the private signing key used by the banknote issuing authority. Furthermore, we assume the adversary is unable to physically clone the same features of a polymer substrate. We emphasise that our security protection is in addition, and orthogonal, to existing security features on a banknote. In reality, a security feature is considered effective if it raises the cost of counterfeiting above the nominal value of the banknote.

### B. Attack on Fingerprints

Like every human being is unique (which forms the basis of "biometrics"), every physical object is unique too (which forms the basis of "physical unclonable function"). Under the microscopic view, every object has distinguished features that can not be exactly duplicated. The same applies to the polymer substrate. Its unevenness in the opacity coating layer reflects the imperfections during the opacity coating, which cannot be avoided. The counterfeiter's challenge is to make another polymer substrate, which gives the same or sufficiently similar feature vector as that of a genuine banknote so the same digital signature can be reused to legitimise the counterfeit.

We argue that it is hard for the attacker to make another polymer substrate that matches a given 2048-bit feature vector even if he has access to the same printing material and equipment as used by the banknote issuing authority. First of all, the attacker needs to produce a "physical" object that looks and feels like a legitimate polymer note. This is substantially harder than launching spoofing attacks in "biometrics", e.g., using a gummy finger to deceive a scanner in an unmanned (unsupervised) environment. By contrast, the verification of banknotes is usually "supervised" by nature. Visual inspection by a human is almost always the first line of defence to detect counterfeits, which is followed by the possible use of tools for further confirmation such as special pens, UV light, or in our case, a film scanner. With reference to the gummy-finger attack, this means an attacker has to make a real "finger" that looks and feels like a human finger to pass the human inspection first, before it can deceive the fingerprint scanner. This is substantially more difficult than a conventional spoofing attack in an unsupervised environment.

The use of the "see-through" window (a security feature of polymer notes) forces the counterfeiter to use a clear plastic film as the substrate. With access to the same printing material and equipment as used by the bank authorities, the attacker will be able to produce polymer substrates that look and feel the same as legitimate notes. However, merely producing another substrate gives only a probability of $p = 5 \times 10^{-25}$ for mismatch, based on an HD threshold of 0.33 (Figure VI). In reality, the feature vector of a second polymer substrate does not have to match exactly that of a target substrate. It only needs to be close enough in the Hamming space, say less than an HD distance of 0.33. Based on the degrees of freedom $N = 900$ (mean HD 0.5), and an HD threshold of 0.33, finding a random N-bit string that is within the $w = 0.33 \cdot N = 297$ bits difference to the target string requires the minimum number of attempts $N'$ as estimated below according to the sphere-packing bound [14].

$$\begin{aligned} N' &= \frac{2^N}{\sum_{i=0}^{w} \binom{N}{i}} \\ &= 4 \times 10^{24} \end{aligned} \tag{15}$$

Note that $N'$ is only a lower bound. The above result implies that if the attacker repeats the same production process, he must produce $4 \times 10^{24}$ polymer substrates in order to find one that might match a given digitally signed fingerprint. This is clearly infeasible for the attacker.

The attacker might improve his chance by adding a custom-built printing step on top of the existing banknote manufacturing process. It is worth noting that printing on a plastic film is much harder than printing on a paper substrate. The novel idea that uses a special plastic film made of BOPP to support high-quality security printing is precisely the key innovation that makes polymer notes possible [3]. However, the film still has to undergo a special opacity coating process to form a polymer substrate, which provides a canvas to allow printing in the subsequent procedure.

As explained earlier, the opacity coating is inherently imperfect, producing a layer of white ink with uneven thicknesses. This leads to random translucent patterns when the light shines thought the substrate. A close-up of the translucent patterns is shown in Figure 10. As shown in the picture, the patterns contain randomly distributed bright spots, as well as dark spots (impurities in the ink). The physical dimensions of these features are on the scale of a few micrometres. As a comparison, high-resolution ink-jet printers use very small drops (normally 17 to 50 pL volume of liquid in one droplet [15]) to create different colours or grey levels. With a volume of $v = 17$ pL, assume it forms a perfect semi-sphere once it falls on the substrate to form a printed dot, the diameter of the dot is $d = 2 \cdot \sqrt[3]{v \cdot 2 \cdot 3/4\pi} = 40\ \mu$m. However, in reality, the droplet collides with the substrate at a high-speed, creating a much larger dot with randomly scattering patterns which resemble nothing like a dot under the microscopic view (e.g., see [16]). While an ink-jet printer is physically limited by the size of the nozzle, an attacker might use a laser printer. However, a laser printer has its own physical limitation. Due to the interaction of multiple rolls, a laser printer prints uncontrollable repeated patterns at the microscopic view [17]. As an experiment, we used two high-resolution inkjet (HP Deskjet 2700) and laser (Kyocera TASKalfa 5052ci) printers to print a dot '.' in different font sizes as shown in Figure 11. The smallest printed size is at least one order of magnitude larger than the size of the impurities observed in the opacity coating layer (see Figure 10). More importantly, the printed dots in Figure 11 exhibit random scattering patterns because the printers cannot precisely control the nozzle or the toner at the microscopic level.

Hence, modern printers have physical limits in what they can print at the microscopic level. While on-top printing can increase the opacity level, the attacker also needs to be able to decrease the opacity level, e.g., by removing white ink in the coating layer, so to have the full control of the translucent patterns. This will require the attacker to acquire much more sophisticated printing equipment than what is used by a legitimate state government. While this is theoretically possible, we believe it is extremely unlikely in practice, and we leave it to further research in the future.

## VIII. Related Work

Vila et al. [18] were among the first to propose analysing the infrared spectrum of a banknote to determine if it is genuine or not. They proposed to examine selected areas of the banknote by using an infrared spectrometer, together with an attenuated total reflectance (ATR) microscope. Their dataset consisted of 18 randomly selected genuine notes of €50 and €100 denominations, and 5 counterfeit notes of €50 and €100 denominations, provided by the Spanish Police. Although their experiments showed distinguishing features in the infrared spectra between the genuine and counterfeit notes, this result critically relied on the specific counterfeit samples used in the study. Sonnex et al. [19] proposed a similar method based on infrared spectroscopy. Their dataset contained 27 counterfeit £20 notes from the Northamptonshire Police. Their study revealed a lack of contrast in infrared spectra between ink and paper among the forgeries. Hence, the authors proposed to use a simple and portable infrared device to search for spectral difference as the first line of defence, and in case of ambiguity, use a more expensive infrared microscope to map selected areas of printing in contrast to the background paper. Their study has the same limitation as [18] in that the result was only applicable to the specific counterfeit samples used in the experiment.

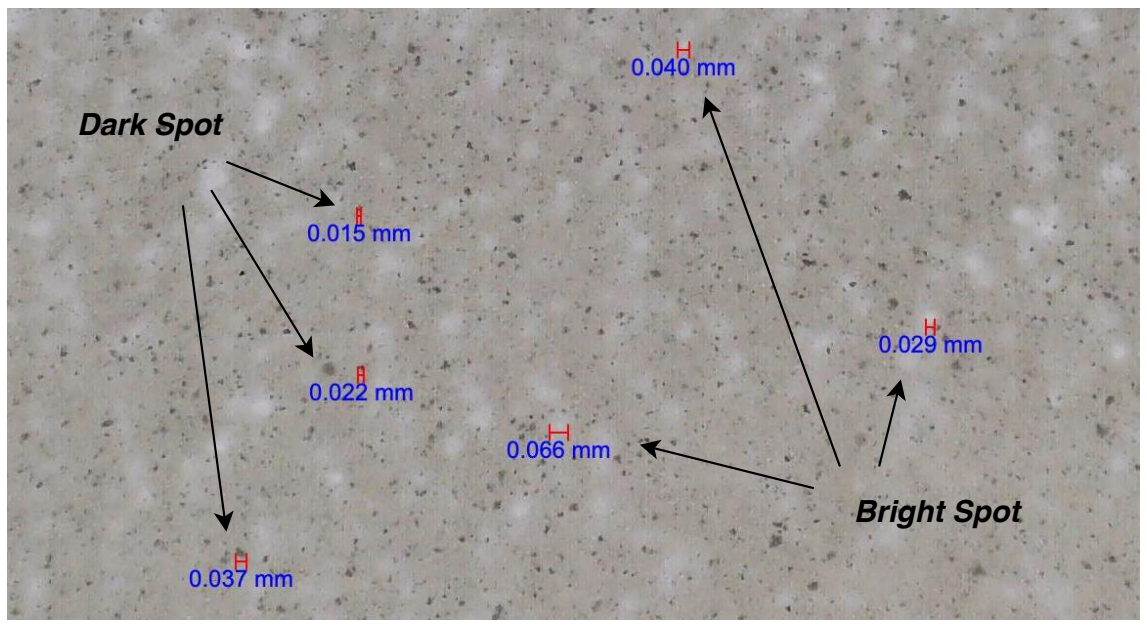

Fig. 10: A close-up of translucent patterns

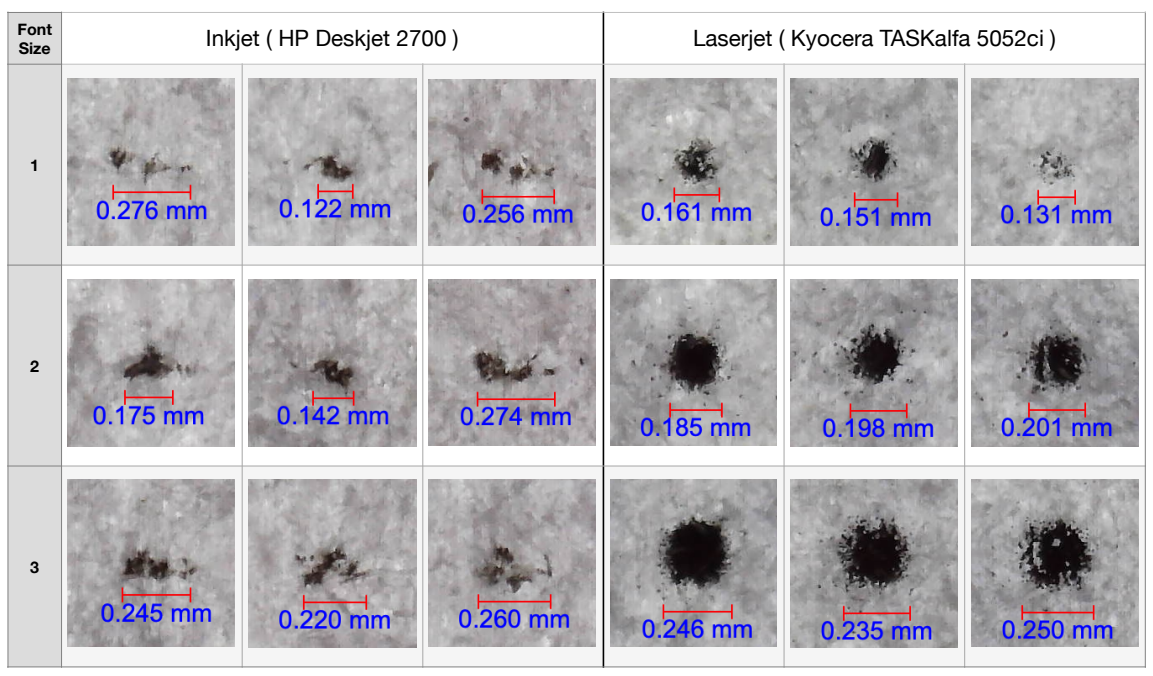

Fig. 11: A printed dot '.' in different font sizes using inkjet and laser printers. The resolution for both printers is 1200 DPI.

Some researchers proposed to analyse the ink composition to distinguish legitimate banknotes from counterfeits. Rusanov et al. [20] applied Mössbauer spectroscopy to analyse the chemical composition of the ink used in both genuine and counterfeit banknotes. They examined 54 authentic $100 US banknotes chosen at random, and 13 forged notes which were provided by a bank in Bulgaria. The authors suggested that the absence of certain elements in the pigment (e.g., green dye sextet) could be used to distinguish counterfeit banknotes from authentic ones. Jara et al. [21] conducted a similar study to analyse the chemical composition of the ink in real and fake banknotes, by using an X-ray fluorescence spectrometer instead. Almeida et al. [22] proposed to apply Raman spectroscopy and chemometric tools to analyse the characterisation of the ink in a banknote. They examined 60 counterfeit banknotes provided by the Brazilian police, and a further set of 28 lab-made fake samples prepared by scanning authentic bills and printing copies on laser and ink-jet printers. Based on the difference in the Raman spectra of the chalcographic ink, the authors proposed to use a Partial Least Square for Discriminant Analysis (PLS-DA) classifier to first distinguish the counterfeits from the originals, and in case of detecting a counterfeit, use a second classier to identify which type of the printer was used in making the counterfeits. The performance of their solution critically depends on the counterfeit samples used in training the classifier. Some researchers proposed to use an imaging device to capture the visual difference between genuine notes and counterfeits. Yeh et al. [23] proposed to analyse the luminance histograms of the captured image of a banknote and apply multiple-kernel support vector machines (SVM) to distinguish the counterfeits from the genuine notes. The authors used a dataset of 70 genuine Taiwanese banknotes and 29 counterfeits. Berenguel et al. [24] proposed a similar technique to detect counterfeits by analysing the background texture printing. A surface picture of a given banknote is taken by using a flatbed scanner and converted to grey-scale. Histogram features of the grey-scale image are extracted as input to a linear SVM classifier to determine if the note is real or not. The authors used a lab-made dataset of forgeries by scanning genuine euro bills and then printed counterfeits with an HP LaserJet printer. In a follow-up paper [25], Berenguel et al. proposed a different classification method, but still used the same procedure to generate lab-made counterfeit samples for evaluation. All of these papers have a common limitation that the results are only valid for the specific counterfeit samples used in the study.

Anti-counterfeiting of banknotes is closely related to anti-counterfeiting of documents, since paper documents such as certificates, cheques and contracts face the same counterfeiting problem as banknotes. Clarkson et al. [16] proposed a method to authenticate a paper document based on the unevenness of its surface. Based on the observation that the fuzz-mat surface of a paper document has a unique 3-D texture structure, they proposed to use a flatbed scanner to scan the target document multiple times at 4 different orientations. Based on the measurements, they created a 3-D image of the paper surface texture, and split the image into small patches for feature extraction. This process created a feature vector of 3200 bits as a paper fingerprint. Experiments showed that their method was able to distinguish a genuine document from a forged one. However, one drawback of their method is that it requires repeated scans, and is time-consuming. In our system, we extract features from a 2D image, and require only one scan (or one snapshot using a camera). Sharma et al. [26] proposed a similar technique based on analysing the speckle patterns when light reflects on the paper surface. This follows an earlier work by Buchanan et al. [27], which used a laser to capture the speckle patterns. Recently, Toreini et al. [11] proposed a new fingerprinting technique, which captures the unique features of a paper document using transmissive light instead of reflective light. They showed that using the transmissive light was able to capture richer features in the textural patterns than using the reflective light, and hence achieve better performance than previous works [16], [26], [27].

Our polymer substrate fingerprinting technique is inspired by the previous research in anti-counterfeiting of banknote and paper, but it is different in a few ways. First of all, we do not require any dataset of forgeries for training. Instead of merely classifying a banknote into a binary result of "real" or "forgery" like in [18]–[25], our technique extracts a unique fingerprint from a physical banknote. The authentication of a banknote starts with a null hypothesis that it is a "forgery" until this hypothesis is compellingly rejected by statistics. The 900-bit entropy in the extracted fingerprints is higher than previous works [11], and lays a solid foundation for building a large-scale authentication system for both online and offline applications. Second, we are the first to propose utilising the imperfections in the opacity coating layer of a polymer banknote to build an anti-counterfeiting system. Besides the theoretical design, we have developed a complete data acquisition and processing method, built a concrete proof-of-concept prototype, collected an extensive dataset and conducted experiments with both empirical and theoretical analysis to demonstrate the feasibility of our proposed solution. As compared with existing security features of polymer notes, ours has a distinctive advantage that even if the attacker has acquired the same printing equipment and ink as used for printing genuine banknotes, and counterfeiting remains hard.

## IX. Conclusion

In this paper, we have proposed a new anti-counterfeiting solution for polymer notes based on analysing the imperfections in the opacity coating of a polymer substrate. The imperfect coating process leaves a coating layer of uneven thickness and randomly distributed impurities from the ink. We propose a method to capture these imperfections and transform them into a 2048-bit feature vector as a unique fingerprint. Our experiments show that our solution is able to authenticate banknotes with high accuracy, is extremely scalable, and is robust against rough daily handling of banknotes. This makes it a useful technique for practical use.